\documentstyle[epsfig,12pt]{article}

\title{
{\bf QUARK-GLUON PLASMA IN THE EARLY UNIVERSE}}
\author{{JOSEPH I. KAPUSTA} \vspace*{0.1in}\\
 {\it School of Physics and Astronomy, University of Minnesota}\\
 {\it Minneapolis, MN 55455, USA}}

\date{}

\begin{document}

\maketitle

\begin{abstract}

A QCD phase transition in the early universe could have left inhomogeneities in 
the baryon to photon ratio and in isospin that might have affected 
nucleosynthesis later on.  At very high temperature QCD plasma can be described 
by perturbation theory because of asymptotic freedom, but a possible phase 
transition requires a nonperturbative approach like lattice gauge theory.  
Assuming that a first order transition did occur, a dynamical set of equations 
can be solved to evolve the universe through it and to quantify the scale of 
inhomogeneity.  Unfortunately this scale appears to be too small by two orders 
of magnitude to affect nucleosynthesis.  

\end{abstract}

\section{Introduction}

Who has not wondered what happens when matter gets squeezed to higher 
and higher densities or heated to higher and higher temperatures?  At first 
molecules would dissociate into atoms, then atoms would ionize, then atomic 
nuclei would undergo a liquid-gas type phase transition to a vapor of protons 
and neutrons.  At high baryon density there will be so many protons and neutrons 
per unit volume that their constituent quarks will no longer know to which 
nucleon they belong.  This is expected to happen when nucleons are about a 
factor of two closer to each other than they are in atomic nuclei, that is, at a 
density of about 8-10 times that in a nucleus.  (Normal nuclear density is 0.15 
nucleons per cubic fermi - equivalent to $2\times10^{14}$ g/cm$^3$.)  At this 
point quarks will be free to move over distances much greater than the 
confinement distance of 1 fm.  This type of matter is a cold quark gas.  It is 
possible, but not yet known, that it exists in the cores of neutron stars.

At high temperature there will be many particle-antiparticle pairs produced.  
The lightest hadrons, the pions, will be the first to be produced in abundance, 
quickly followed by all the hadrons listed in the Particle Data Tables 
\cite{PDG}.  Hagedorn realized already in the 1960's that the level density of 
hadrons grows exponentially $dN/dm \sim \exp(m/T_0)$ where $m$ is the hadron 
mass \cite{Hag}.  The Hagedorn temperature $T_0$ has the numerical value 160 
MeV.  A thermal ensemble of hadrons treated as noninteracting particles 
therefore has a limiting temperature equal to $T_0$ because of this exponential 
growth.  In a sense it is the end of the hadronic world; higher temperatures 
bring a quark-gluon plasma.  Intuitively, this transition comes about because 
the hadron density is so great that the constituent quarks and gluons are 
confused about which hadron they ought to be confined to.

The main interest in a cosmological hadron to quark-gluon plasma phase 
transition arises from its potential to influence big bang nucleosynthesis, 
discussed in section 2.  Whether or not QCD with its known set of parameters 
undergoes a first order transition or something smoother is still not settled.  
This issue is discussed in section 3 using perturbation theory and lattice 
gauge theory.  Assuming that there is a first order phase transition one needs 
nucleation theory to understand how the transition proceeds, and this topic is 
discussed in section 4.  The cosmological QCD phase transition is then analyzed 
in section 5.  Conclusions, and paths for future work, are given in section 6.

\section{Inhomogeneous Big Bang Nucleosynthesis}

A cosmological first order phase transition at $T \sim 160-180$ MeV could leave 
spatial inhomogeneities in the baryon to entropy ratio and in the ratio of 
protons to neutrons.  If these inhomogeneities survive to $T \sim 0.1-1$ MeV 
they could influence nucleosynthesis.  This was first pointed out and analyzed 
by Witten \cite{W}, by Applegate, Hogan, and Scherer \cite{AHS}, and by Alcock, 
Fuller, and Mathews \cite{AFM}.  In thermal and chemical equilibrium one might 
expect that the baryon density in the quark-gluon phase is higher than in the 
hadron phase.  This is called the baryon density contrast.  Assuming a critical 
temperature of $160 < T_c < 180$ MeV, Olive and I computed this baryon density 
contrast to be 1.5 to 2.5 when hadronic interactions were neglected and 5 to 7 
when they were included \cite{KO}.  One would expect the last regions of space 
to undergo the phase conversion to contain more baryons per unit volume than the 
first regions to phase convert because of the lack of time for baryons to 
diffuse.  After phase completion the neutrons will diffuse more rapidly than 
protons because they are electrically neutral and therefore do not Coulomb 
scatter on electrons.  This leads to isospin inhomegeneities, at least 
temporarily.

A detailed calculation of inhomogeneous nucleosynthesis with a comparison to 
observed abundances of the light elements was done by Kurki-Suonio, Matzner, 
Olive, and Schramm \cite{KMOS}.  They considered baryon density contrasts 
ranging from 1 to 100 and a fraction of matter in the high density regions 
ranging from 1/64 to 1/4.  The average separation of the high density regions 
$l$ was left as a free parameter as was the average baryon to photon ratio of 
the universe.  Differential diffusion of protons and neutrons was accounted for 
and then a standard nucleosynthesis code was run.  By fitting the observed 
abundances of $^4$He, D, $^3$He, and $^7$Li they concluded that the baryon to 
photon ratio must lie between $2\times10^{-10}$ and $7\times10^{-10}$ (or 
$2\times10^{-10}$ if certain constraints on $^7$Li were relaxed).  They also 
concluded that $l < 150$ m at the time of nucleosynthesis.  At the completion of 
the QCD phase transition this upper limit would have been about 1 m.  Early 
estimates of this distance were uncertain but promising \cite{early1,early2}.
See also the reviews listed in \cite{reviews}.  A better estimate of this scale
is the purpose of sections 4 and 5.

Recently the inhomogeneous nucleosynthesis calculation was redone with technical 
improvements and updated estimates of the cosmic abundances of the relevant 
light elements by Kainulainen, Kurki-Suonio, and Sihvola \cite{KKS}.  Their 
results are shown in figure 1.
\begin{figure}
\centerline{\epsfig{figure=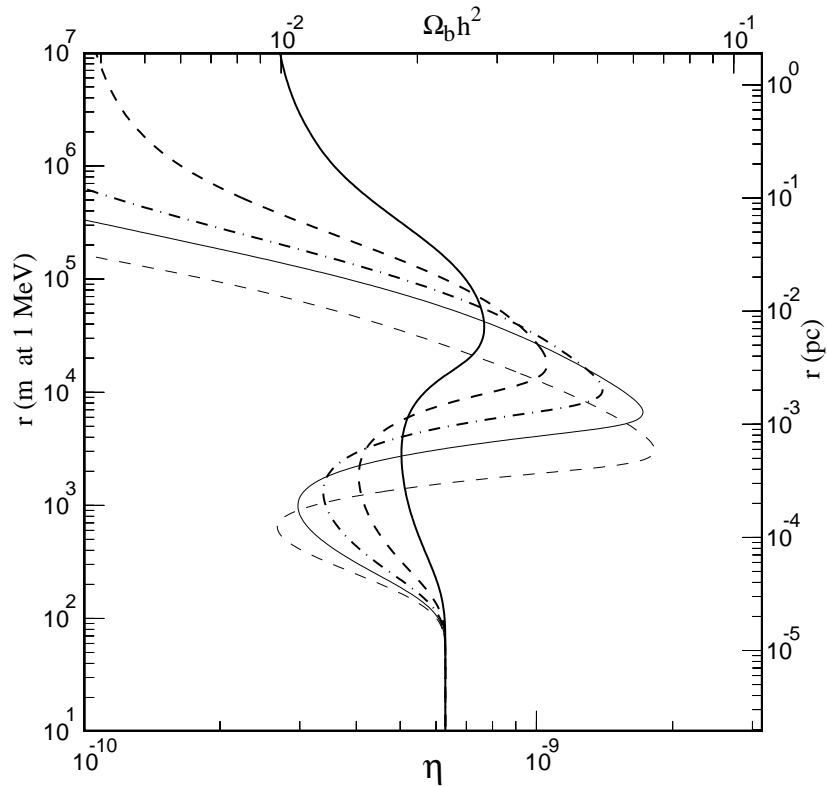,width=11.0cm}}
\caption{Conservative upper limit to the baryon to photon ratio $\eta$
from the $^4$He abundance $Y_p \leq 0.248$ and the deuterium abundance
D/H$\geq 1.5\times10^{-5}$.  The thick curves are for volume fractions
covered by the high density regions of $1/(2\sqrt{2})$ (solid), $1/8$
(dashed), and $1/(16\sqrt{2})$ (dot-dashed).  The thin curves are for
$1/64$ (solid) and $1/256$ (dashed).  From \cite{KKS}.}
\end{figure}
The high density matter was distributed in spheres.  The 
inhomogeneities are ineffective in influencing nucleosynthesis unless the high 
density regions were separated by more than about 150 m at $T = 1$ MeV.
   
\section{QCD Equation of State}

The QCD equation of state can, in principle, be computed from the 
functional integral formula for the partition function \cite{kapbook}.
\begin{equation}
Z = \int [d\overline{\psi}][d\psi][dA^a_{\mu}][d\overline{C}_a]
[dC_a] \exp\left\{\int_0^{\beta}d\tau \int_V d^3x \, {\cal L}_{eff}
\right\}
\end{equation}
Here $\psi$ is a quark field, $A^a_{\mu}$ is the gluon field, and $C_a$ is the 
ghost field.  The Lagrangian is
\begin{eqnarray}
{\cal L}_{eff} &=& {\cal L} -\frac{1}{\rho} \left(\partial^{\mu}
A_{\mu}^a \right)^2 +\mu_f \overline{\psi}_f \gamma^0 \psi_f
+\partial_{\mu}\overline{C}_a \partial^{\mu}C_a
+g f_{abc}\overline{C}_a\partial_{\mu}A^{\mu}_b C_c \, ,\\ 
{\cal L} &=& -\frac{1}{4} F^{\mu\nu}_a F_{\mu\nu}^a
-\overline{\psi}_f \left(i\not\!\partial -m_f - 
g\not \!\!A^a G^a\right) \psi_f \, ,\\
F^{\mu\nu}_a &=& \partial^{\mu}A^{\nu}_a - \partial^{\nu}A^{\mu}_a
-g f_{abc} A^{\mu}_b A^{\nu}_c \, .
\end{eqnarray}
The subscript $f$ refers to quark flavor. The $G^a$ are the generators of the 
color group $SU(N_c)$.  The quark fields are Grassmann 
variables antiperiodic in imaginary time $\tau = it$, $0 < \tau < \beta
= 1/T$, the gluon fields are periodic in this time, and the ghost fields are 
Grassmann variables periodic in imaginary time.  (The complex ghost field may be 
thought of as subtracting 2 of the 4 degrees of freedom implied by $A^{\mu}$ to 
yield the 2 physical transverse degrees of freedom of the massless gluon.)  The 
Lagrangian is written in covariant gauge with $\rho$ the gauge parameter.
Physical observables are independent of $\rho$.

There are two well-known ways to evaluate this functional integral.  One is 
perturbation theory in the coupling constant $g$; the other is numerical 
estimation using Monte Carlo techniques on a discrete lattice.  Before reviewing 
results obtained in these approaches, let us try to predict what the result 
might be.  Suppose that all quarks were massless and that all chemical 
potentials were zero.  Then there is only one parameter in the QCD Lagrangian, 
the coupling $g$ which is dimensionless.  If this coupling is really a constant 
then dimensional analysis implies that the pressure must be some function of g 
times the temperature to the fourth power: $P = f(g)T^4$.  Then the entropy 
density is $s = 4f(g)T^3$, and the equation of state may be written as $\epsilon 
= 3P$, the same as a free gas of massless particles.  This is a trivial equation 
of state without a phase transition.  In contrast the quantum version of the 
theory displays dimensional transmutation.  The dimensionless coupling constant 
$g$ gets replaced by a renormalization group running coupling.  At finite 
temperature and at one-loop order \cite{CP}
\begin{equation}
g^2 \rightarrow \overline{g}^2 = \frac{24\pi^2}{(11N-2N_f)
\ln\left(T/\Lambda_{QCD}\right)}
\end{equation}
for an $SU(N_c)$ gauge group ($N_c = 3$ for QCD) and $N_f$ flavors of quark.  
The scale parameter $\Lambda_{QCD}$, which has dimensions of energy, enters as a 
result of the necessity to regulate and renormalize, that is, to render 
correlation functions finite at some arbitrarily chosen energy or momentum 
point.  As a result the equation of state with $P = f(\overline{g}(T))T^4$ can 
now become complicated and not predictable by dimensional analysis alone.  This 
allows nontrivial thermodynamics and the possibility of a phase transition.  
This is on account of quantum mechanics, not classical interacting field theory.

The partition function for QCD was calculated within perturbation theory to 
order $g^3$ in the late 1970's by myself and then to order $g^5$ in the mid 
1990's by Arnold and Zhai \cite{AZ}, by Kastening and Zhai \cite{KZ}, and by 
Braaten and Nieto \cite{BN}.  For $N_f$ flavors of massless quarks and zero 
chemical potentials the pressure is
\begin{eqnarray}
P &=& \frac{8\pi^2}{45}T^4 \left[ 1 + \frac{21}{32}N_f -
\frac{15}{4} \left(1+\frac{5}{12} N_f \right)
\frac{\alpha_s(M)}{\pi} \right. \nonumber \\
&+& 30\left(1+\frac{1}{6}N_f\right)^{3/2} 
\left(\frac{\alpha_s(M)}{\pi}\right)^{3/2} +
F_4\left(N_f,\ln(M/2\pi T)\right) \left(\frac{\alpha_s(M)}{\pi}\right)^2
\nonumber \\
&+& \left. F_5\left( N_f,\ln(M/2\pi T)\right) \left(\frac{\alpha_s(M)}
{\pi}\right)^{5/2} + {\cal O}\left( \alpha_s^3 \ln\alpha_s \right)
\right] \, .
\end{eqnarray}
Here $\alpha_s = g^2/4\pi$, and $F_4$ and $F_5$ are lengthy untransparent 
functions of the number of flavors and of the renormalization scale $M$.  The 
latter should be chosen judiciously so as to minimize the contribution of higher 
order terms in the expansion.  For sake of illustration consider $N_f=3$, 
$M=2\pi T$, $T = 500$ MeV, and $\alpha_s \approx 0.2$.  Then 
\begin{equation}
P = \frac{19\pi^2}{36}T^4 \left[ 1 - 0.18 + 0.30 + 0.06 - 0.37 ... \right]
\end{equation}
representing the contributions of $\alpha_s$ to the power of 0, 1, 3/2, 2, and 
5/2.  At so low a temperature, only about three times the expected $T_c$, this 
expansion is apparently not very quickly converging.  Although very useful for 
very high temperature because asymptotic freedom forces the running coupling to 
zero logarithmically, perturbation theory cannot reliably predict the location 
or order of a possible phase transition.

By its very nature perturbation theory using free particle basis states is 
incapable of describing a phase transition.  An alternative is to discretize 
space and (imaginary) time and estimate the functional integral expression for 
$Z$ numerically with Monte Carlo techniques \cite{MC1}.  One works with a 
spatial volume $V=(N_sa)^3$ and a time dimension $\beta=N_{\tau}a$, where $a$ is 
the lattice spacing and $N_s$ and $N_{\tau}$ are the number of sites in each 
direction.  The thermodynamic limit is $a\rightarrow 0$, $N_s \rightarrow 
\infty$, $N_{\tau} \rightarrow \infty$, subject to $\beta$ held fixed and $V 
\rightarrow \infty$.  The difficulties associated in reaching for the 
thermodynamic limit 
include the enormous computer power and storage necessary for the huge number of 
degrees of freedom inherent in a quantum field theory, and the theoretical 
difficulty with describing fermions on a lattice.  A typical lattice for pure 
gauge theory with no quarks is $64^3\times 8$, and with quarks is $16^3\times 
4$.  These are not very large lattices, especially in the latter case.  
Nevertheless computer power and storage is increasing with time, and 
practitioners are constantly improving the numerical algorithms.  For example, 
improved lattice actions, where derivatives are better approximated via higher 
order finite differences and short distance behavior can be supplemented with 
perturbation theory on account of asymptotic freedom, have helped significantly.
Here I have relied on a recent review by Karsch \cite{Karsch}.
    
Pure quarkless $SU(2)$ gauge theory is expected to undergo a second order phase 
transition reflecting color deconfinement at high temperature whereas $SU(N_c)$ 
with $N_c>2$ is expected to undergo a first order phase transition.  The $SU(3)$ 
equation of state computed in lattice gauge theory is shown in figure 2.
\begin{figure}
\centerline{\epsfig{figure=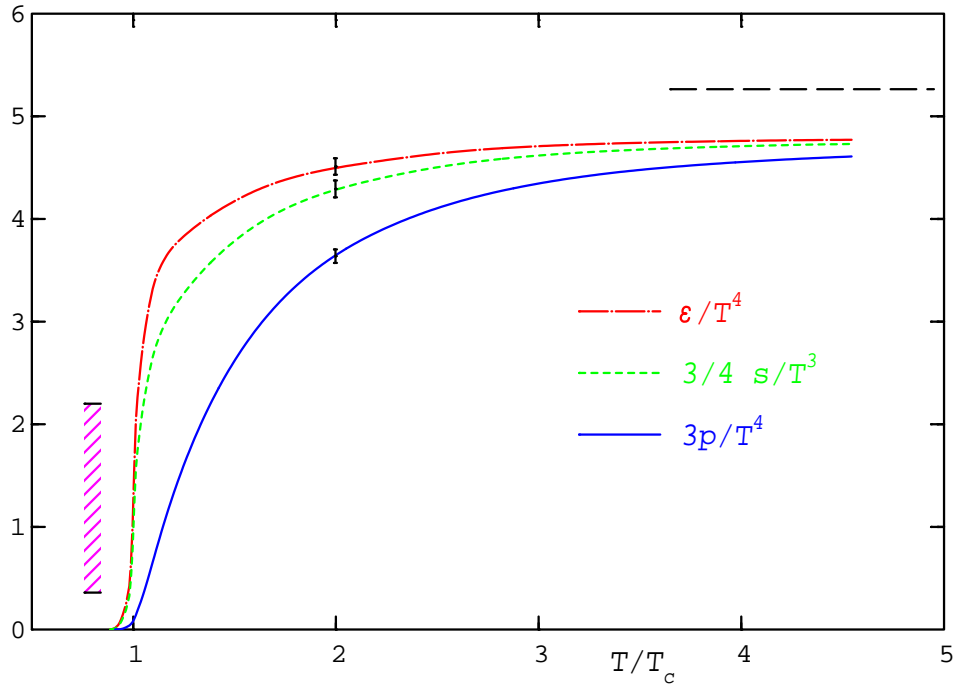,width=13.5cm}}
\caption{The equation of state of pure $SU(3)$ gauge theory with no quarks.
From \cite{KKS}.}
\end{figure}
One physical quantity is required to set the overall energy scale.
This is chosen 
to be the string tension $\sigma$ at zero temperature; its numerical value is 
taken to be the same as in our physical world.  Then the critical temperature is 
about 270 MeV and the latent heat is about 0.5 GeV/fm$^3$.  Below $T_c$ is a gas 
of (presumably) weakly interacting very massive glueballs.  The lightest one is 
1700 MeV.  Above $T_c$ all 16 degrees of freedom of massless (interacting) 
gluons are liberated.  These computational results seem to be stable with time 
and therefore well established and accepted.

\newpage

A compendium of results from different groups for QCD with two degenerate 
flavors of quarks is shown in figure 3.
\begin{figure}
\centerline{\epsfig{figure=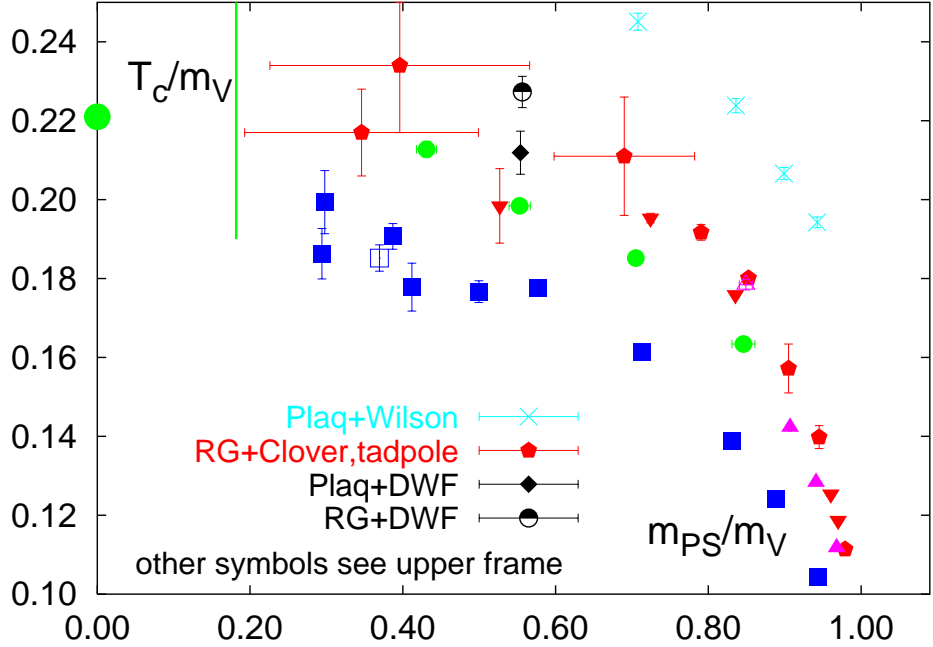,width=13.5cm}}
\caption{$T_c$ as a function of the lightest pseudoscalar meson (pion)
with both measured in units of the lightest vector meson mass ($\rho$ or
$\omega$) calculated by several groups with a variety of lattice
methods.   From \cite{Karsch}.}
\end{figure}
In the limit that the quarks are 
massless a second order transition is expected, with no proper thermodynamic 
transition when they are massive.  Even with massive quarks one sees a sharp 
rise in the energy density in a narrow window of temperature. The middle of this 
range can be used to define a crossover temperature still called, somewhat 
incorrectly, $T_c$.  In order to set the physical energy scale the mass of the 
lightest vector meson, the $\rho$-meson, is calculated and set equal to its real 
world value of 770 MeV.  The mass of the lightest pseudoscalar meson, 
the $\pi$-meson, is calculated and compared to the real world value of 
140 MeV.  The pseudoscalar mass is a monotonically increasing function of the 
quark mass so it can be used to determine what quark mass ought to be used in 
the lattice calculation to represent as closely as possible the real world.  The 
$T_c$ is plotted against the ratio of the pseudoscalar to vector meson masses, 
in units of the vector mass.
Taking the limit of zero quark mass is notoriously difficult.  The 
actual range of computed pion masses starts at 1.2 GeV and decreases to 300 MeV 
but no lower.  There is a lot of scatter among the results from different groups 
using different methods but the trend is towards $T_c = 170$ MeV in the zero 
quark mass limit.

\newpage

QCD with three flavors of massless quarks is expected to undergo a first order 
phase transition.  The question is: What is the order of the transition for the 
real world with up and down quark masses of order 5-7 MeV and a strange quark 
mass of order 110-150 MeV?  Some results are shown in figure 4.
\begin{figure}
\centerline{\epsfig{figure=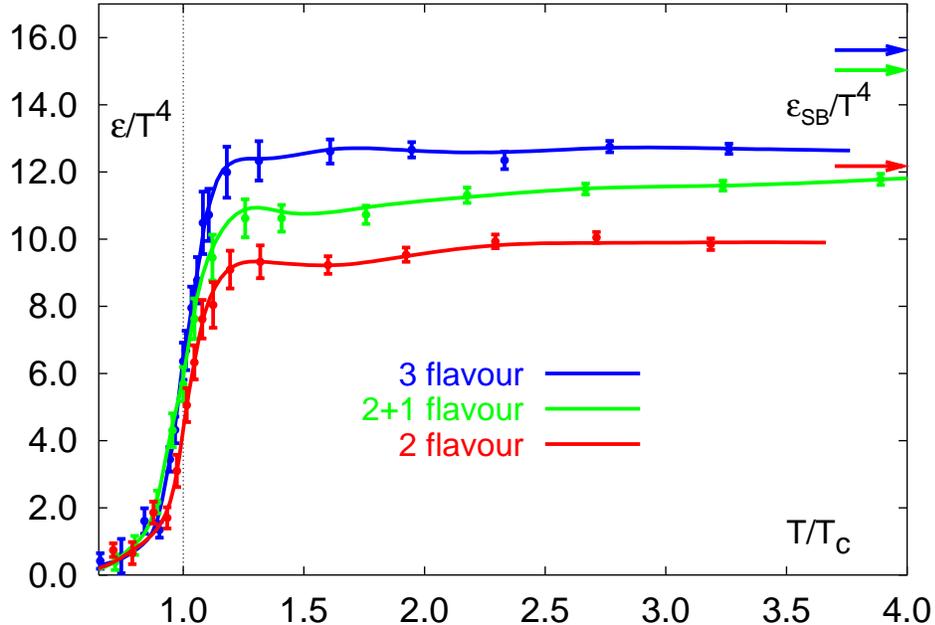,width=13.0cm}}
\caption{Energy density in units of $T^4$ as a function of $T$ for
two light quarks (2 flavor), two light quarks and one heavy quark (2+1 flavor),
and three light quarks (3 flavor). From \cite{Karsch}.}
\end{figure}
The lattice is 
$16^3\times 4$ in size.  The quark masses are actually temperature-dependent 
(the usual situation for numerical reasons) with $m_{\rm light} = 0.4T$ and 
$m_{\rm heavy} = T$.  Results for two light flavors, two light and one heavy 
flavor, and three light flavors are shown in this figure.  The energy
density $\epsilon$, in units of $T^4$, is plotted against $T/T_c$.  In all 
three cases there is a big jump in the energy density over a narrow window of 
temperature.  The 2+1 flavor case is almost indistinguishable from the 3 flavor 
case in the critical region.  The statistical error bars are very small.  More 
significant are the systematic errors that are much more difficult to estimate.  
Due to the relatively small lattice size and still rather large light quark mass 
it is difficult to say whether or not the real world has a first order phase 
transition.  Even if not, the large jump in energy density may make the real 
situation practically the same as if it did.         

\section{Relativistic Nucleation Theory}

The dynamics of first-order phase transitions has fascinated scientists
at least since the time of Maxwell and Van der Waals.  Much work on the
classical theory of nucleation of gases and liquids was carried out in
the early part of the 20th  century.  There were and still
are many important applications, such as cloud and bubble chambers,
freezing of liquids, and precipitation in the atmosphere. 
The goal of nucleation theory is to compute the probability that a bubble
or droplet of one phase appears in a system initially in the other phase
near the critical temperature.  Homogeneous nucleation theory applies
when the system is pure; inhomogeneous nucleation theory applies
when impurities cause the formation of bubbles or droplets.  For the
application we have in mind, namely the early universe,
it seems that homogeneous nucleation theory
is appropriate.  In the everyday world it is usually the opposite; dust
or ions in the atmosphere are much more efficient in producing precipitation.
Nucleation theory is applicable for first-order phase transitions when
the matter is not dramatically supercooled or superheated.  If substantial
supercooling or superheating is present, or if the phase transition is
second-order, then the relevant dynamics is spinodal decomposition.

Suppose that a system is cooled below its critical temperature.  Then there
exists a critical sized droplet (or bubble, depending on whether the energy
density in the lower temperature phase is greater or less than the higher
temperature phase).  If a droplet that forms because of statistical
fluctuations is too small, its surface free energy is relatively large
and the cost in total free energy is positive.  The droplet will evaporate.
If the droplet is large, its surface free energy is unimportant, and the
droplet will accrete molecules and grow.  A droplet of critical size
is metastable, it is balanced between evaporation and accretion.  The
classical theory of Becker and D\"oring \cite{BeD35}, which is nicely
reviewed by McDonald \cite{McD}, says that the probability per unit time
per unit volume to nucleate the dense liquid phase from a dilute
gas is given by
\begin{equation}
I = a(i_*) \left(\frac{\epsilon''(i_*)}{2\pi T}\right)^{1/2}
\bar{n}(1) e^{-\epsilon(i_*)/T} \, ,
\end{equation}
where $\epsilon(i_*)$ is the formation energy of a critical sized droplet
consisting of $i_*$ molecules, prime denotes differentiation with respect
to the number of molecules $i$, $\bar{n}(1)$ is the density of single molecules 
and $a(i_*)$ is the accretion rate of single molecules on a critical droplet.  
Usually the accretion rate is taken to be
\begin{equation}
a(i_*) \,=\, \frac{1}{2} \bar{n}(1) \bar{v} 4\pi r_*^2 s \, ,
\end{equation}
which is the flux of particles ($\bar{v}$ is the mean speed of gas
molecules) striking the surface of the critical droplet times
a `sticking fraction' $s$ less than one.  The first
term in this formula is a dynamical factor influencing the growth rate, the
second term characterizes fluctuations about the critical droplet, and
the product of the third and fourth terms gives the quasi-equilibrium
number density of critical sized droplets.

The modern theory of nucleation was pioneered by
Langer \cite{Lang}.  Langer's theory is based in a
more fundamental way on the microscopic interactions of atoms and
molecules.  It can also be applied close to a critical point where in fact
most of the current interest in the condensed matter community has been.
Langer's modern theory of nucleation yields the following formula for
the rate:
\begin{equation}
I = \frac{\kappa}{2\pi} \Omega_0 e^{-\Delta F_*/T}
\end{equation}
where $\Delta F_*$ is the change in the free energy of the system due to the
formation of the critical droplet. $\Omega_0$ is a statistical prefactor
which measures the available phase space volume.
\begin{equation}
\Omega_0 = 2 \left( \frac{\sigma}{3T}\right)^{3/2}
\left(\frac{r_*}{\xi_g}\right)^4
\end{equation}
Here $\xi_g$ is the correlation length in the gaseous phase.
$\kappa$ is a dynamical
prefactor which determines the exponential growth rate of critical droplets
which are perturbed from their quasi-equilibrium radius $r_*$.
\begin{equation}
\kappa = \frac{d}{dt}\ln [r(t) - r_*]
\end{equation}
The basic structure is the same
as in the classical theory, but the prefactors are different.
The dynamical prefactor has been calculated by Langer and Turski
\cite{LaT80,LaT73} and by Kawasaki \cite{Kawa}
for a liquid-gas phase transition near the critical point, where
the gas is not dilute, to be
\begin{equation}
\kappa = \frac{2\lambda\sigma T}{\ell^2 n_{\ell}^2 r_*^3} \, .
\end{equation}
This involves the thermal conductivity $\lambda$, the surface free energy
$\sigma$, the latent heat per molecule $\ell$ and the density of molecules
in the liquid phase $n_{\ell}$.
The interesting physics in this expression is the appearance
of the thermal conductivity.  In order for the droplet to grow beyond
the critical size, latent heat must be conducted away from the surface into
the gas.

For a relativistic system of particles or quantum fields which
has no net conserved charge, such as baryon number, the thermal conductivity
vanishes.  The reason is that there is no rest frame defined by the baryon
density to refer to heat transport.  Hence the Langer formula cannot
be applied to such systems.  The extension to such systems was made by
Csernai and me \cite{CK1}.  The change in free energy due to the appearance of a 
bubble of hadronic matter in quark-gluon plasma is
\begin{equation}
\Delta F = \frac{4\pi}{3} r^3 \left[P_{\rm q}(T)-P_{\rm h}(T)\right]
+ 4\pi r^2 \sigma \, ,
\end{equation}
where $r$ is the radius.  The critical sized bubble has radius
\begin{equation}
r_* = \frac{2\sigma}{P_{\rm h}(T)-P_{\rm q}(T)}
\end{equation}
which leads to
\begin{equation}
\Delta F_* = \frac{4\pi}{3} \sigma r_*^2 \, .
\end{equation}
The expression for the statistical prefactor is unchanged, but the formula for 
the dynamical prefactor, or growth rate, is proportional to the shear $\eta_q$ 
and bulk $\zeta_q$ viscosities in the quark-gluon plasma instead of the thermal 
conductivity.
\begin{equation}
\kappa = \frac{4\sigma(3\zeta_q+4\eta_q)}{3(\Delta w)^2 r_*^3}
\end{equation}
This is inversely proportional to the square of the enthalpy ($w=\epsilon+P)$ 
difference between the two phases.  The nucleation rate is
\begin{equation}
I = \frac{4}{\pi} \left( \frac{\sigma}{3T}\right)^{3/2}
\frac{\sigma(3\zeta_q+4\eta_q)r_*}{3(\Delta w)^2 \xi_q^4}
e^{-\Delta F_*/T} \, .
\end{equation}
For numerical purposes we use an MIT bag model type equation of state with
\begin{eqnarray}
P_q &=& \left(45.5+14.25\right) \frac{\pi^2}{90}T^4 - B \nonumber \\
P_h &=& \left(5.5+14.25\right) \frac{\pi^2}{90}T^4 \, .
\end{eqnarray}
The 45.5 approximates the effective number of degrees of freedom arising from 
massless gluons and up and down quarks and a strange quark mass comparable to 
the temperature.  The 5.5 approximates the hadronic equation of state near $T_c$ 
arising from a multitude of massive hadrons.  The 14.25 arises from photons, 
neutrinos, electrons, and muons common to both phases.  The bag constant $B$ is 
chosen to give $T_c = 160$ MeV.  Furthermore Csernai and I \cite{CK2} estimated that $\sigma = 50$ MeV/fm$^3$, $\xi_q = 0.7$ fm, and $\eta_q = 18 T^3$ \cite{baym}.  Generally for relativistic systems the bulk viscosity is small in comparison to the shear viscosity and so we neglect it.

\section{Cosmological QCD Phase Transition}

Given the nucleation rate one would like to know the (volume) fraction of
space $h(t)$ which has been converted from the quark-gluon plasma to the 
hadronic gas at the proper time $t$ in the early universe.
This requires kinetic equations that use
the nucleation rate $I$ as an input. 
Here we use a rate equation first proposed by Csernai and me \cite{CK2}. The 
nucleation rate $I$ is the probability to form a bubble of critical
size per unit time per unit volume.  If the system cools to $T_c$
at time $t_c$ then at some later time $t$ the fraction
of space which has been converted to the hadronic phase is
\begin{equation}
h(t) = \int_{t_c}^t dt' I(T(t')) [1-h(t')] V(t',t).
\end{equation}
$V(t',t)$ is the volume of a hadronic bubble at time $t$ which was
nucleated at the earlier time $t'$; this takes into account bubble growth.
The factor $1-h(t')$ takes into account the fact that new bubbles
can only be nucleated in the fraction of space not already occupied
by the hadronic gas.  This conservative approach neglects any spatial variation 
in the temperature.  However, it is an improvement over the formula proposed by 
Guth and Weinberg \cite{GW} for first order cosmological phase transitions in 
that it allows for completion of the transition: the Guth and Weinberg formula 
is only valid during the early stage of the transition.

Next we need a dynamical equation which couples the time evolution
of the temperature to the fraction of space converted to the hadronic phase.  The following analysis is very similar to the one used to study the electroweak phase transition in the early universe by Carrington and me \cite{tau}.
We use Einstein's equations as applied to the early universe,
neglecting curvature.  The evolution of the energy density is
\begin{equation}
\frac{d\epsilon}{dR} = -\frac{3w}{R} \ ,
\end{equation}
where $R$ is the scale factor at time $t$.
This assumes kinetic but not phase equilibrium, and is basically a
statement of energy conservation.  We express the energy density as
\begin{equation}
\epsilon = h\epsilon_h(T) + [1-h]\epsilon_q(T),
\end{equation}
where $\epsilon_h$ and $\epsilon_q$ are the energy densities in the two phases 
at the temperature $T$.  There is a similar equation for the enthalpy $w$.
The time dependence of the scale factor is determined by the equation
of motion
\begin{equation}
\frac{1}{R} \frac{dR}{dt} = \sqrt{\frac{8\pi G \epsilon}{3}} \ .
\end{equation}
This expression can be used to relate the time to the scale factor
using the normalization $R(t_c) = 1$.

We also need to know how fast a bubble expands once it is created.  This
is a subtle issue since by definition a critical size bubble is metastable
and will not grow without a perturbation. 
After applying a perturbation, a critical size bubble begins to
grow.  As the radius increases, the surface curvature decreases, and an
asymptotic interfacial velocity is approached.  The asymptotic radial growth
velocity will be referred to as $v(T)$.
The expected qualitative behavior of $v(T)$ is that the closer $T$ is
to $T_c$ the slower the bubbles grow.  At $T_c$ there is no motivation
for bubbles to grow at all since one phase is as good as the other.  The bubble 
growth velocity was studied by Miller and Pantano \cite{MP}.  Their 
hydrodynamical results may be parameterized by the simple formula
\begin{equation}
v\gamma = 3\left(1-\frac{T}{T_c}\right)^{3/2}
\end{equation}
which indeed has the expected behavior.
A simple illustrative model for bubble growth is then
\begin{equation}
V(t',t) = \frac{4\pi}{3}\left( r_*(T(t')) + \int_{t'}^t dt'' v(T(t''))
\right)^3 \, .
\end{equation}
This expression can also be written in terms of $R, R', R''$ instead of
$t, t', t''$.

We now have a complete set of coupled integro-differential equations which must 
be solved numerically.  These equation take into account bubble nucleation and 
growth, energy conservation, and Einstein's equations.  They make no assumption 
about entropy conservation. 

Figure 5 shows the temperature as a function of the scale factor.
\begin{figure}
\centerline{\epsfig{figure=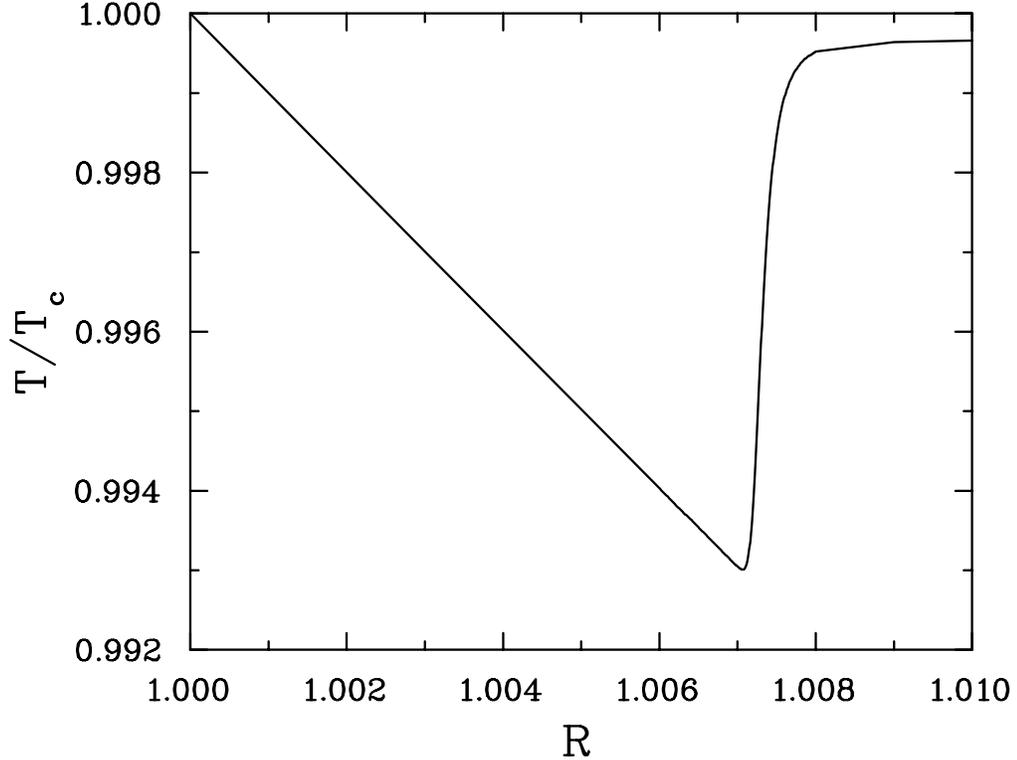,width=10.0cm,angle=90}}
\caption{Temperature as a function of scale factor.}
\end{figure}
For practical purposes, nucleation begins near the bottom of the
cooling line.  Thereafter, nucleation and growth of bubbles
releases latent heat, which causes the temperature to rise.  The increasing 
temperature shuts off nucleation, and the phase transition continues due to the 
growth of already nucleated bubbles.  The temperature can never quite reach 
$T_c$; if it did, bubble growth would cease and the transition would never 
complete.  This is a result of the equations of motion and is not an imposition. 

Figure 6 shows the average bubble density
\begin{equation}
n(R(t)) = \int_{t_c}^t dt' I(T(t')) [1-h(t')]
\end{equation}
as a function of the scale factor. The bubble density rises rapidly just before 
$R$ reaches 1.007 and reaches its asymptotic value just after 1.007.
\begin{figure}
\centerline{\epsfig{figure=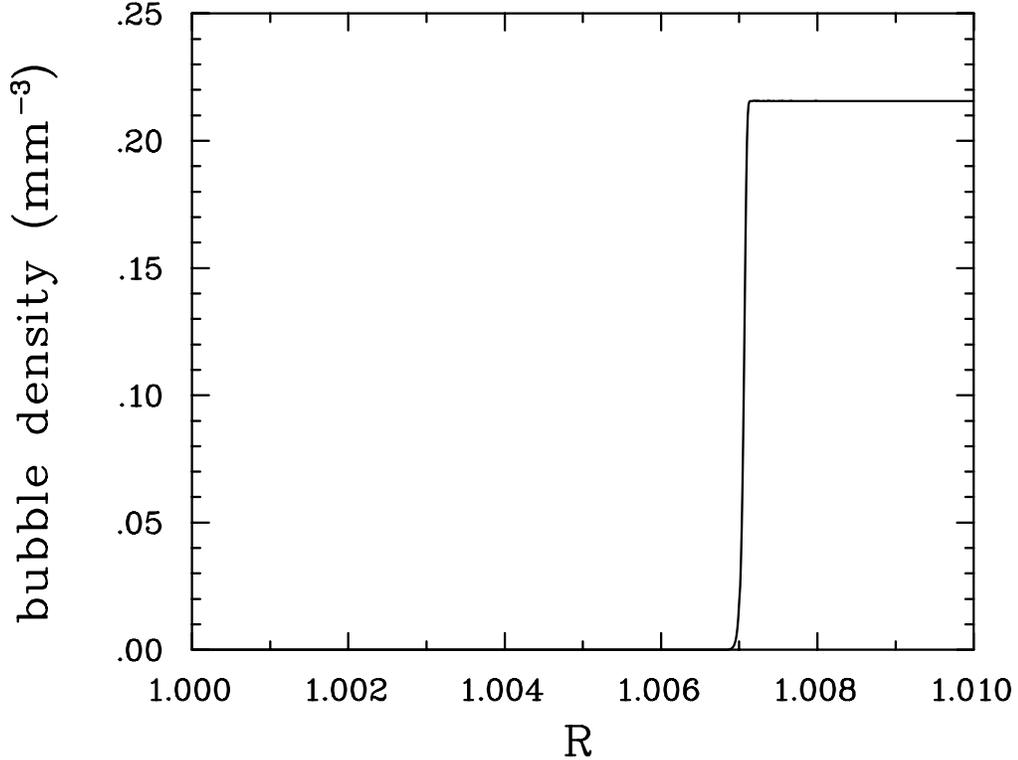,width=10.0cm,angle=90}}
\caption{Average bubble density as a function of scale factor.}
\end{figure}

Figure 7 shows the nucleation rate as a function of scale factor.  The rate has 
a very sharp maximum between 1.0070 and 1.0071.  The turn on and turn off of the 
nucleation rate corresponds precisely with the fall and rise of the temperature 
shown in figure 5.

Figure 8 shows the fraction of space $h$ which has made the
conversion to the hadronic phase.  When $h=1$ the transition is complete and the 
temperature will begin to fall again.  This occurs when $R_f \approx 1.4464$, to 
be compared with the value one would obtain from an ideal Maxwell construction 
$R_{\rm Maxwell} = (239/79)^{1/3} = 1.44630...$.
In fact the whole curve $h(R)$ is very 
close to the ideal Maxwell construction, apart from its delayed start apparent 
in the figure.  The interested student will work out the Maxwell formula from 
the equations given here.

\begin{figure}
\centerline{\epsfig{figure=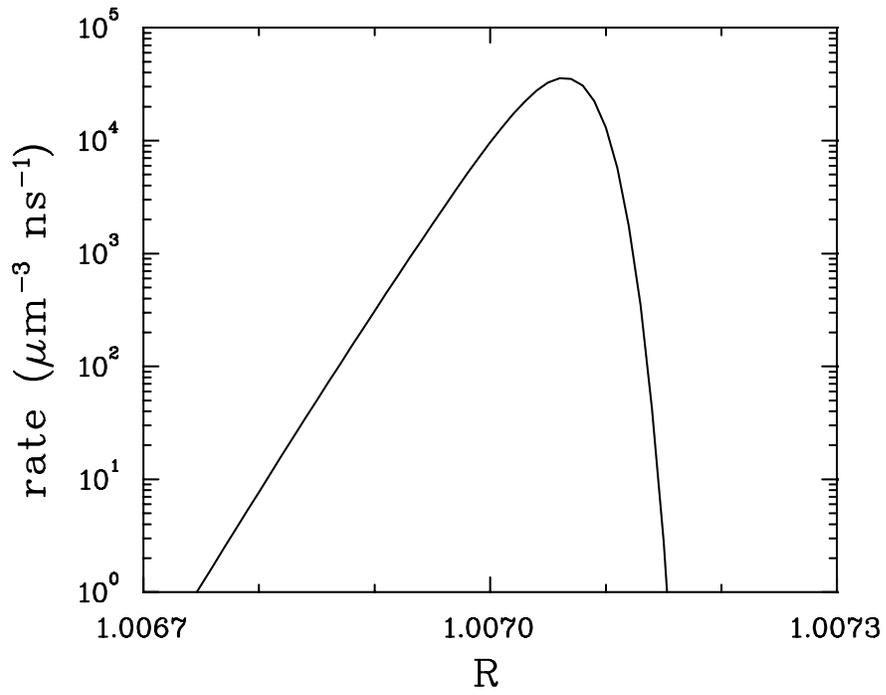,width=9.0cm,angle=90}}
\caption{Nucleation rate as a function of scale factor.}
\end{figure}
\begin{figure}
\centerline{\epsfig{figure=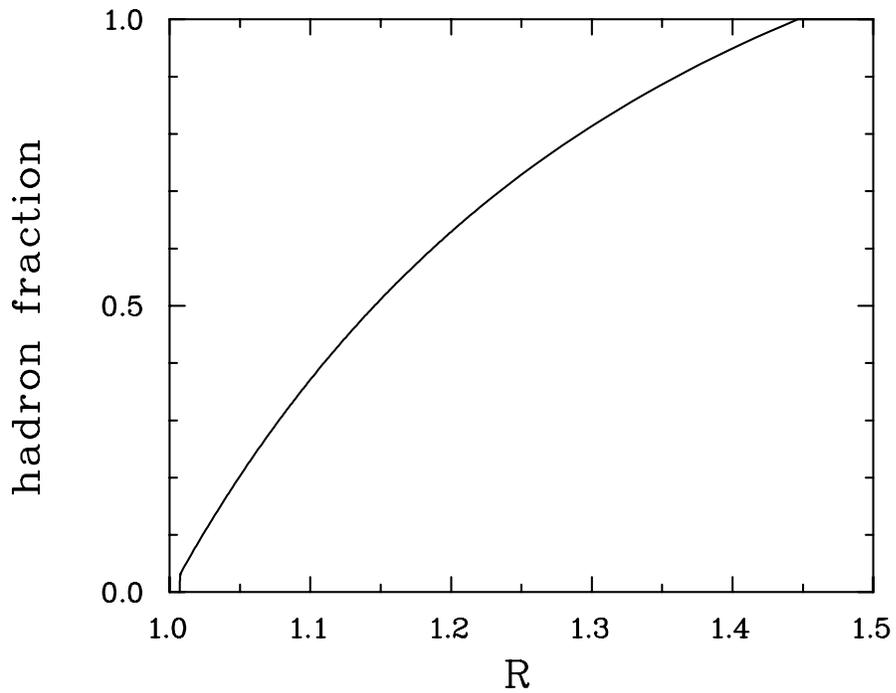,width=9.0cm,angle=90}}
\caption{Volume fraction of space occupied by the hadronic phase as a
function of scale factor.}
\end{figure}

Figure 9 shows the average bubble radius as a function of scale factor.
\begin{equation}
\frac{4\pi}{3}\overline{r}^3 \, n = h
\end{equation}
It grows with time and with the scale factor, of course.  At the end of the 
phase transition it is of the order of 1 cm.  This is also the order of 
magnitude of the distance between the final quark-gluon plasma regions.  
Unfortunately, this is two orders of magnitude too small to affect 
nucleosynthesis.  This result is rather robust against reasonable variations in 
any of the input parameters.
\begin{figure}
\centerline{\epsfig{figure=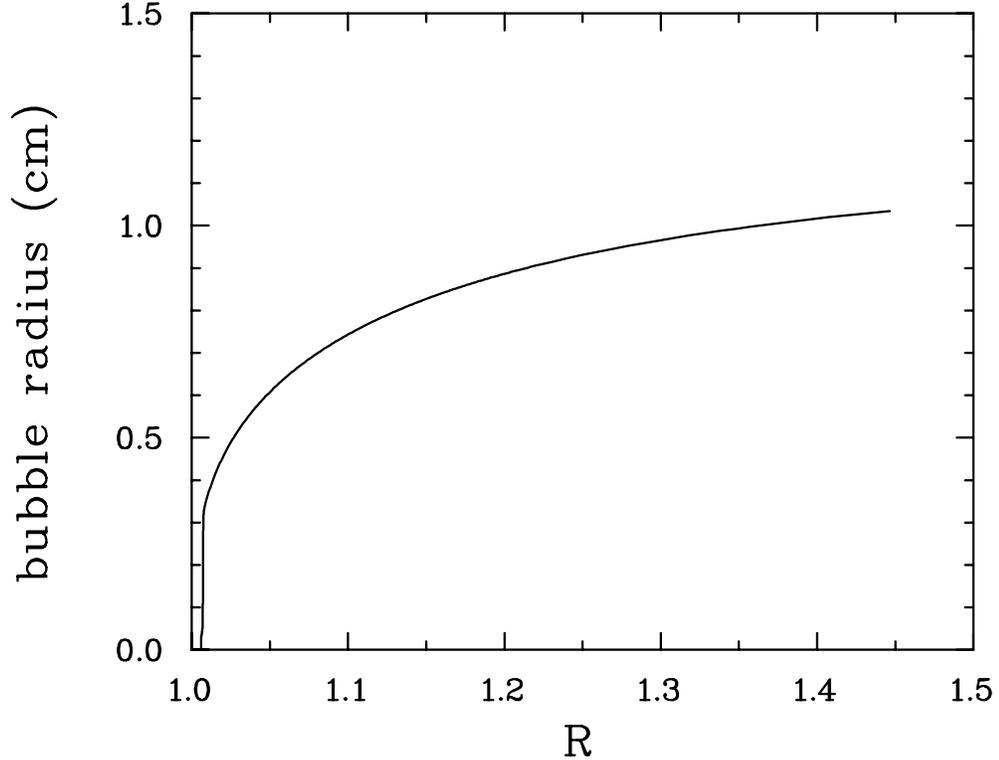,width=10.0cm,angle=90}}
\caption{Average bubble radius as a function of scale factor.}
\end{figure}   

\section{Conclusion}
 
At temperatures exceeding approximately 160-180 MeV matter takes the form of quark-gluon plasma rather than hadronic gas.  In principle QCD can be used to calculate the partition function and all other properties of interest.  Perturbation theory has been employed to calculate the equation of state at asymptotically high temperatures to order $g^5 \sim \alpha_s^{5/2}$, but the results diverge at low temperature and cannot predict the properties of a phase transition.  For this purpose numerical Monte Carlo simulations of lattice gauge theory are necessary.  For the real world with 2 light flavors and one moderately light flavor of quark there is a big increase of the energy density in a narrow window of temperature centered on 170 MeV.  Despite twenty years of effort it is still not known whether there is a true thermodynamic phase transition of first or second order or just a very rapid crossover.  This is due to the great difficulty of computation with light dynamical quarks on the lattice.

Nucleosynthesis is affected by remnant inhomogeneities in the baryon to entropy ratio and isospin if the high baryon density regions immediately following a QCD phase transition are separated by at least 1 m.  A set of dynamical equations can be written and solved for the evolution of the universe through such a phase transition all the way to completion.  The evolution of the temperature and hadronic volume fraction as functions of time and scale factor are hardly any different than the results of an idealized Maxwell construction.  The information not available in the latter construction is the length scale of the inhomogeneities, that is, bubble sizes and so on.  The characteristic distance between the last regions of quark-gluon plasma seem to be on the order of 1 cm, too small to affect nucleosynthesis.  However, there are qualifications and improvements can be made.  For example, when the fraction of space
occupied by bubbles exceeds about 50\%, interactions among the
bubbles probably cannot be neglected.  It is unlikely, though,
that further improvements in the dynamics would qualitatively
change the current picture of the transition.  Indeed, crude
estimates of the effects of bubble fusion on the dynamics of the
QCD transition in heavy ion collisions indicate that the transition completes 
only a little faster, and that the average bubble size is greater
\cite{Kluge}.  At least that is in the right direction to be interesting.

The Relativistic Heavy Ion Collider (RHIC) is an accelerator at Brookhaven National Laboratory designed and constructed explicitly to search for 
quark-gluon plasma and the nature of a possible phase transition.  Operations began in the summer of 2000, and a wealth of data is expected soon.  Essentially RHIC is attempting to reproduce the first microsecond of the early universe, to produce matter with temperatures exceeding well over 200 MeV.  Many of us look forward to exploitation of this little big bang in the laboratory to cosmology \cite{qm}. 

\section*{Acknowledgements}

This work was supported by the US Department of Energy under grant
DE-FG02-87ER40328.

\end{document}